\documentstyle[aps]{revtex}
\begin{document}
\draft
\title{
Stars and Halos of Degenerate Relativistic
Heavy-Neutrino and Neutralino Matter
}
\author{
Neven Bili\'c$^{1,2}$,
Faustin Munyaneza$^1$,
and
Raoul D.~ Viollier$^1$
}
\address{
$^1$Department of Physics,
University of Cape Town, Rondebosch 7700, South Africa
\\
 $^2$Rudjer Bo\v{s}kovi\'{c} Institute, 10000 Zagreb, Croatia
}
\date{\today}
\maketitle
\begin{abstract}
Heavy-neutrino (or neutralino) stars are
studied using the general relativistic equations
of hydrostatic equilibrium and the relativistic equation
of state for degenerate fermionic matter.
The Tolman-Oppenheimer-Volkoff equations
are then generalized
 to include a
system of degenerate neutrino and neutralino matter
that is gravitationally coupled.
The properties and implications of such an interacting
astrophysical system are discussed in detail.
\end{abstract}

\section{Introduction}
\label{sec1}
One of the most tantalizing puzzles of this universe is
the issue of
 dark
matter, the presence of which is
inferred from the observed flat rotation curves in
spiral galaxies \cite{FABER,TRIMBLE}, the
diffuse emission of x-rays in elliptical
galaxies and clusters of galaxies,
 as well as from cluster  dynamics.
Primordial nucleosynthesis entails that most of
 baryonic matter in this universe is nonluminous,
 and such an amount
 of dark matter falls suspiciously
close to that required by  galactic rotation curves.
However,
although a significant component of  dark matter
in galactic halos is presumably baryonic \cite{COTU}, the
bulk part of  dark matter in this universe
is believed to be nonbaryonic.
Many candidates have been proposed \cite{PEEB}, both
  baryonic as well as nonbaryonic,
 to explain the dark matter paradigm,
 but the issue of the nature of dark matter
 is still far from being resolved.

One of the most conservative candidates for nonbaryonic
dark matter are, of course, massive neutrinos.
In this paper
we are particularly interested in
 neutrinos with masses
between 10 and  25 keV/$c^{2}$,
as these could form supermassive degenerate neutrino stars,
which may explain, without invoking
 the black-hole hypothesis, some
 of the  features
observed  around the supermassive
 compact dark  objects
with masses ranging from
$10^{6.5}~M_{\odot}$ to $10^{9.5}~M_{\odot}$\,.
These have been reported to exist
at the center of a number of
galaxies \cite{KORI,KOB,MYM,mac} including
 our own \cite{GETO,ESGE,MO,BTV,TV,TVR} and
quasistellar objects.
It is interesting to note that neutrinos in this mass
range can also cluster around ordinary stars,
and thus these neutrinos could account for
at least part of  galactic dark matter.
A further motivation for studying the collapsed
structures of heavy neutrino matter
 is the recent increased interest
in fermionic cold dark matter  models \cite{NAF}
in which massive
neutrinos  play an important
role in structure formation
in the  early universe.

A 10 to 25 keV/$c^{2}$ neutrino is in conflict
neither
with particle and nuclear physics experiments nor with
astrophysical observations \cite{RDV}.
On the contrary, if the conclusion of the LSND collaboration
which claims to have
 detected $\bar{\nu}_{\mu} \rightarrow \bar{\nu}_{e}$
flavor oscillations \cite{ATH} is confirmed, and
the quadratic see-saw mechanism involving
the up, charm, and top quarks \cite{GERA,YANA}, is the correct
mechanism for neutrino mass generation,
the $\nu_{\tau}$  mass may  be
 between 6 and 32 keV/$c^{2}$ \cite{BVI},
 which is well
 within the cosmologically
forbidden range.
It is  well known that such a quasistable neutrino would lead
to an early neutrino-matter dominated phase,
which may have started as early as a couple of weeks
after the big bang.
Thus, a critical universe that
remained  neutrino-matter dominated
all the time
  would have reached the current microwave
background temperature
in less than 1 Gyr,
i.e., much too early
to accommodate the oldest stars in globular clusters,
cosmochronology, and the Hubble expansion age.

It is conceivable, however, that, in the presence of such
heavy neutrinos, the early universe
might have evolved quite differently than
described in the homogeneous standard model of cosmology.
Neutrino stars may have emerged in local
condensation processes during a gravitational phase transition,
shortly after the neutrino-matter dominated epoch began.
The latent heat produced in such a first-order phase transition,
apart from reheating the gaseous phase, might have reheated
the radiation background as well.
Annihilation of  heavy neutrinos into
light neutrinos via the $Z^{0}$
would take place in the interior
of neutrino stars \cite{Lurie,RDV}.
 Both these processes would decrease
the density of heavy neutrinos,
as perceived
today, and also
increase the time which photons need
to cool down to the present microwave background
temperature.
Thus a quasistable neutrino in the mass range between
10 and 25 keV/$c^{2}$
is presumably not in contradiction with
cosmological and
astrophysical observations \cite{RDV}.

In fact, it has recently been shown \cite{BTV,BVI,BIVI} that
degenerate neutrino stars \cite{RDV,Lurie,VILT,VTT}
may  indeed have been formed during
 a gravitational phase transition in the early universe.
Whereas the existence of this first-order phase
transition is firmly established in the framework
of the Thomas-Fermi model
 at finite temperature \cite{BIVI},
the microscopic mechanism through which the latent heat
 is released during
 the phase transition and
 dissipated into observable and perhaps unobservable matter
or radiation remains to be identified.
At this
stage, however, it is still not clear whether
an efficient dissipation mechanism
can be found within the minimal extension of the
standard model of particle physics or whether
 new physics is required in the right-handed neutrino
sector.
We therefore have to assume in the following
that such an efficient dissipation mechanism
exists, in order to make sure that fermions
can actually settle in the state of lowest energy
in a time much shorter than the age of the
universe.

In this paper, we focus
primarily  on
gravitationally clustered, degenerate nonbaryonic
 matter consisting of two species
of weakly interacting stable or quasistable fermions:
one with a mass around 15  keV/$c^{2}$ which we
subsequently call  ``neutrino" , and the other
with a mass around 1 GeV/$c^{2}$ which we henceforth call
``neutralino".
The chosen neutralino mass,
although a little bit on the low side,
offers the possibility of replacing the neutralino with a
neutron,
as the strong-interaction effects of the neutron
in neutron  star matter  can be
simulated by an effective mass.
 Of course, this substitution
 makes sense only as long as
 the binding energy of the neutron
is larger  than the Q value for the neutron decay, so that the
neutron can be considered stable in neutron star matter.

It is interesting to note that
a variety of similar scenarios  can be
treated within the same framework: Apart from a neutrino halo
around a neutron star, one could also study a neutrino halo
around a white dwarf or
 around an ordinary star \cite{VILT},
since all these baryonic stars can be approximated
using
similar polytropic equations of state which
 eventually result in the same
nonlinear differential equations of the
Lan\'e-Emden type.
Moreover, by varying the polytropic index of the equation
of state, one can also investigate the properties of a cold
 neutrino star immersed in a hot radiation field,
or in a hot baryonic background, or in a vacuum with
nonzero energy density, which all may have played a role in the
formation process of primordial neutrino stars.
Thus the study of this simple interacting
neutrino-neutralino
system allows us to learn a great deal about
the properties  of gravitationally clustered
baryonic and nonbaryonic matter.

This paper is organized as follows:
In Sec.~\ref{sec2}
in a general relativistic framework,
we discuss
the properties and implications of
 degenerate neutrino (and neutralino) stars
and their Newtonian and ultrarelativistic
limits.
In Sec.~\ref{sec3} we generalize
the Tolman-Oppenheimer-Volkoff (TOV) equations to
 include gravitationally clustered,
degenerate  nonbaryonic
matter, consisting of neutrinos and neutralinos.
  Our results are summarized
in Sec.~\ref{sec4}.

\section{Degenerate  neutrino stars}
\label{sec2}

 A spherically symmetric cloud of degenerate neutrino
matter can be characterized
 by its mass density $\rho_{\nu}(r)$,
 pressure $P_{\nu}(r)$, and the metric in the
 Schwarzschild form \cite{ZE}
\begin{equation}
ds^{2}={\rm{e}}^{\nu}c^{2}dt^{2}
-{\rm{e}}^{\lambda}dr^{2}-r^{2}(d\theta^{2}
+\sin^{2}\theta d\phi^{2}) .
\label{eq:01}
\end{equation}
The pressure and the density
 satisfy the general relativistic TOV
 equations
 of hydrostatic equilibrium \cite{TO,OV}:
\begin{equation}
\frac{dP_{\nu}}{dr}=-\frac{1}{2}
(\rho_{\nu}c^{2}+P_{\nu})\frac{d\nu}{dr},
\label{eq:02}
\end{equation}
\begin{equation}
{\rm{e}}^{\lambda}=\left(1-\frac{2Gm}{c^{2}r}\right)^{-1},
\label{eq:03}
\end{equation}
\begin{equation}
\frac{dP_{\nu}}{dr}=-G
\frac{(\rho_{\nu}+P_{\nu}/c^2)
(m+4\pi r^{3}P_{\nu}/c^2)}{r(r-2Gm/c^{2})},
\label{eq:04}
\end{equation}
\begin{equation}
\frac{dm}{dr}=4\pi r^{2}\rho_{\nu}(r) ,
\label{eq:05}
\end{equation}
where $m(r)$ is the mass enclosed  within a radius $r$.
The relevant boundary conditions are
$\ m(0)=0$, $P_{\nu}(R)=0$, and $\rho_{\nu}(R)=0$, as the pressure
and the density
vanish at the radius $R$ of the star.
Outside the star,  the functions
$\nu$ and $\lambda$ are determined by
 the usual Schwarzschild solution
\begin{equation}
{\rm{e}}^{\nu}={\rm{e}}^{-\lambda},\;\;\;\;\;
 \ {\rm{e}}^{\lambda}=(1-2GM/c^{2}r)^{-1},
\label{eq:06}
\end{equation}
\begin{equation}
M=\int_{0}^{R}4\pi\rho_{\nu}(r)r^{2}dr .
\label{eq:07}
\end{equation}
We now introduce the  equation of state of a degenerate
  relativistic Fermi gas \cite{TEU}
which  may be  parameterized as
\begin{equation}
P_{\nu}=K\left[X(1+X^{2})^{1/2}\left(\frac{2}{3}X^{2}-1\right)+
\log\left(X+(1+X^{2})^{1/2}\right)\right],
\label{eq:08}
\end{equation}
\begin{equation}
\rho_{\nu}=\frac{K}{c^2}\left[X(1+X^{2})^{1/2}(2X^{2}+1)
-\log\left(X+(1+X^{2})^{1/2}\right)\right],
\label{eq:09}
\end{equation}
\begin{equation}
n_{\nu}=\frac{8KX^3}{3m_{\nu}c^{2}}\  .
\label{eq:10}
\end{equation}
Here,
$n_{\nu}$ denotes the neutrino-number
 density, and  $K$ and
 $X$  are given by
\begin{equation}
K=\frac{g_{\nu}m_{\nu}c^5}{16\pi^{2}\hbar^{3}},
\;\;\;\;  X=\frac{p_{\nu}}{m_{\nu}c},
\label{eq:11}
\end{equation}
where
$p_{\nu}$ stands for the local Fermi
 momentum of the neutrinos of mass $m_{\nu}$,
and $g_{\nu}$ is the
spin degeneracy factor of
 neutrinos and antineutrinos, i.e., $g_{\nu}=2$
 for Majorana and
$g_{\nu}=4$ for Dirac neutrinos and antineutrinos.
Using  (\ref{eq:08}) and (\ref{eq:09}),
and
introducing  dimensionless variables $x=r/a_{\nu}$
 and $\mu= m/b_{\nu}$
with the scales
\begin{equation}
a_{\nu}=
2 \sqrt{\frac{\pi}{g_{\nu}}}\left(
\frac{M_{\rm Pl}}{m_{\nu}}\right)^2 L_{\rm Pl}
=
2.88233\times10^{10}g_{\nu}^{-1/2}
\left(\frac{17.2\ {\rm keV}}{m_{\nu}c^{2}}\right)^{2}{\rm km},
\label{eq:12}
\end{equation}
\begin{equation}
b_{\nu}=
2 \sqrt{\frac{\pi}{g_{\nu}}}\left(
\frac{M_{\rm Pl}}{m_{\nu}}\right)^2 M_{\rm Pl}
=1.95197\times10^{10}M_{\odot}g_{\nu}^{-1/2}
\left(\frac{17.2\ {\rm keV}}{m_{\nu}c^{2}}\right)^{2} ,
\label{eq:13}
\end{equation}
where $M_{\rm{Pl}}=(\hbar c/G)^{1/2}$
 and  $L_{\rm{Pl}}=(\hbar G/c^3)^{1/2}$ denote
  Planck's  mass and length, respectively,
  the
 TOV equations (\ref{eq:04}) and
(\ref{eq:05}) can be written as
\begin{equation}
\frac{dX}{dx}=-\frac{1+X^{2}}{X(x^{2}-2\mu x)}
\left\{\mu+x^{3}
\left[X(1+X^{2})^{1/2}\left(\frac{2}{3}X^{2}-1\right)
 +\log\left(X+(1+X^{2})^{1/2}\right)\right]\right\},
\label{eq:14}
\end{equation}
\begin{equation}
\frac{d\mu}{dx}=x^{2}\left[X(1+X^{2})^{1/2}(2X^{2}+1)
-\log\left(X+(1+X^{2})^{1/2}\right)\right] ,
\label{eq:15}
\end{equation}
subject to the boundary conditions $X(0)=X_{0}$ and $\mu(0)=0$.
 In addition to
(\ref{eq:14}) and (\ref{eq:15}), there is also  an equation
governing the number of neutrinos $n$
   within a radius $r=a_{\nu}x$:
\begin{equation}
\frac{d\tilde{n}}{dx}=x^{2}X^{3}(1-2\mu/x)^{-1/2}\ ,
\label{eq:16}
\end{equation}
where $\tilde{n}=n/N_{0}$
is the rescaled neutrino-number density
subject to the boundary condition $\tilde{n}(0)=0$,
with
\begin{equation}
N_{0}=\frac{8b_{\nu}}{3m_{\nu}}=3.3765\times10^{72}
\left(\frac{17.2\ {\rm keV}}{m_{\nu}c^{2}}\right)^{3}g_{\nu}^{-1/2}.
\end{equation}

 Equations (\ref{eq:14})-(\ref{eq:16}) may be
 solved numerically.
Picking up a value $X_{0}$
for the Fermi momentum at the center
(in units of $m_{\nu}c$),
one obtains
the total mass of the star
$M$, the radius $R$, and the total number of particles $N$,
 by integrating
 outward until
$X$ vanishes.
The results are summarized in Figs. \ref{fig1}
and \ref{fig2}.
In Fig.~\ref{fig1}
the total mass is plotted
 against the radius of the neutrino  star.
The curve has a maximum, namely, the
 Oppenheimer-Volkov (OV) limit
\cite{OV},
 at $\mu_{\rm{OV}}=0.15329$, which  corresponds
to a neutrino star mass of
\begin{equation}
M_{\rm{OV}}=0.15329 \, b_{\nu}
  =
 0.54195\, M_{\rm{Pl}}^{3}m_{\nu}^{-2}g_{\nu}^{-1/2}
   =
  2.9924\times10^{9}M_{\odot}
\left(\frac{17.2\ {\rm keV}}{m_{\nu}c^{2}}\right)^{2}g_{\nu}^{-1/2}.
\label{eq:mm}
\end{equation}

Owing to their large mass, neutrino stars could serve
as candidates for
 supermassive compact
  dark objects
 observed
in the mass range
\begin{equation}
2.5\times10^{6}~M_{\odot}\   \stackrel{\textstyle <}{\sim}\  M
 \stackrel{\textstyle <}{\sim} 3\times10^{9}~M_{\odot}
\label{eq:00}
\end{equation}
at the centers of a number of galaxies.
Assuming that the most massive
and violent objects
are neutrino stars
 at the OV limit with
$M_{\rm{OV}}=(3.2\pm 0.9) \times 10^{9}~M_{\odot}$,
such as the supermassive compact dark object
at the center of M87
\cite{mac},
 the neutrino mass
 required for this scenario is
\begin{eqnarray}
 12.4\,{\rm keV}/c^2
 \leq &m_{\nu}&  \leq
 16.5\,{\rm keV}/c^2
 \;\;\;\;\; {\rm for}
\;      g_{\nu}=2,
\nonumber \\
 10.4\,{\rm keV}/c^2
 \leq &m_{\nu}&  \leq
 13.9\,{\rm keV}/c^2
 \;\;\;\;\; {\rm for}
\;      g_{\nu}=4.
\label{eq:1100}
\end{eqnarray}
The radius of such a neutrino star  is
$R_{\rm{OV}}=4.4466~R_{\rm{OV}}^{s}$,
 where $R_{\rm{OV}}^{s}=2G M_{\rm{OV}}/c^2$ is the
Schwarzschild radius of the mass $M_{\rm{OV}}$.
Thus, at a distance of a few Schwarzschild radii away from the
 supermassive object,
there is little difference between a neutrino star
at the OV limit
and a black hole, in particular
since the last stable orbit around a black hole
already has a radius of $3~R_{\rm{OV}}^{s}$.
A neutrino star of mass
$M_{\rm{OV}}=3\times 10^{9}~M_{\odot}$
would have a radius
$R_{\rm{OV}}=3.9396\times 10^{10}$ km,
or 1.52 light-days.

Of course, neutrino stars that  are
well below the OV limit  will have a size much
larger than black holes of the same mass,
although they will still be dark and much
more compact than any known baryonic
object of the same mass.
As the gravitational potential
of such an extended neutrino star
is much shallower, significantly
less energy will be dissipated through
accreting matter than in the case of a black hole
of the same mass.
In fact, there is compact dark matter at the
center of our galaxy with
$(2.45\pm0.40)\times 10^6
~M_{\odot}$
concentrated within a radius smaller than
0.0254 pc or 30.3 light-days   \cite{GETO,ESGE},
determined from the motion of stars
in the vicinity of Sgr A$^*$.
Interpreting this supermassive compact
dark object in terms of a degenerate neutrino star
of $2.5\times10^6
~M_{\odot}$,
the upper limit for the size of the object
provides us with a lower limit for
the neutrino mass, i.e.,
\begin{eqnarray}
  &m_{\nu}&  \geq
 14.3\,{\rm keV}/c^2
 \;\;\;\;\; {\rm for}
\;      g_{\nu}=2,
\nonumber \\
  &m_{\nu}&  \geq
 12.0\,{\rm keV}/c^2
 \;\;\;\;\; {\rm for}
\;      g_{\nu}=4.
\label{eq:1200}
\end{eqnarray}
In this context,
it is important to note that
if Sgr A$^*$ is a matter-accreting neutrino star
\cite{BTV,TV,TVR},
one can, in a natural way, explain the so-called
``blackness problem" of
 Sgr A$^*$ ,
 i.e., the fact that
 Sgr A$^*$  does not seem to emit
 detectable x-rays of a few tens of keV,
 which would be emitted by baryonic matter
 falling towards a black hole.
 As this unmistakable black-hole
 signature is missing,
 the concept of a
 ``black hole on starvation"
 has been created in order to save
 the black-hole idea.
 However, the neutrino-star model also fits
 the enigmatic radio-emission spectrum of
 Sgr A$^*$  much better than the
 ``black hole on starvation"
 model
 \cite{TVR}.

The total mass of the neutrino star $M$
is plotted against the total
number of particles $N$ in Fig.~\ref{fig2}.
For masses much smaller than the OV limit, the relation
between $M$ and $N$ is unique.
However, as $M$ approaches the
OV limit, $M$ becomes a multivalued function of $N$.
The part of the curve on the left side of the maximum in
Fig.~\ref{fig1},
 which corresponds to the upper part of the
curve in Fig.~\ref{fig2}, represents unstable
configurations \cite{ZE,HAT} for which the relative mass defect
\begin{equation}
\Delta=\frac{Nm_{\nu}-M}{Nm_{\nu}}
\label{eq:17}
\end{equation}
eventually becomes negative, as seen in
Fig.~\ref{fig3}.
Thus, for $\Delta < 0$, the system can gain energy
by disintegrating.
The maximal relative mass defect,
or the strongest binding,
is obtained at the OV limit with
$\Delta_{\rm{OV}}=3.5807\times 10^{-2}$.

For completeness, we note
that  in the Newtonian limit  $X_{0} \ll 1$,
the TOV  equations (\ref{eq:14}) and (\ref{eq:15})
reduce  to
\begin{equation}
\frac{dX}{dx}=-\frac{\mu}{x^{2}X},
\label{eq:18}
\end{equation}
\begin{equation}
\frac{d\mu}{dx}=\frac{8}{3}x^{2}X^{3}\ ,
\label{eq:19}
\end{equation}
which, using the
 substitution $\Theta=X^{2}$  and $\xi=4x/\sqrt{3}$,
can be cast into the nonlinear Lan\'e-Emden differential
equation with the polytropic
 index $3/2$ \cite{CHANDRA}
\begin{equation}
\frac{1}{\xi^2}\frac{d}{d\xi}\left(\xi^{2}
\frac{d\Theta}{d\xi}\right)=-\Theta^{3/2}\  .
\label{eq:21}
\end{equation}
Owing to the scaling property of the Lan\'e-Emden equation,
 the mass and radius of a nonrelativistic neutrino star
scale as \cite{RDV}
\begin{equation}
MR^{3}=\frac{91.869\, \hbar^{6}}{G^{3}m_{\nu}^{8}}
\left(\frac{2}{g_{\nu}}\right)^2.
\label{eq:0001}
\end{equation}
In the limit $X_{0} \ll 1$, (\ref{eq:08}) and (\ref{eq:09})
yield the
 equation of state of a  nonrelativistic
degenerate Fermi gas, i.e.,
\begin{equation}
P_{\nu}=\left(\frac{6}{g_{\nu}}\right)^{2/3}\rho_{\nu}^{5/3}
\frac{\pi^{4/3}\hbar^{2}}{5m_{\nu}^{8/3}},
\end{equation}
as expected.

For large central densities $ X_{0} \gg 1$,
 $\mu$  oscillates around  $\mu_{\infty}=0.09196$,
which corresponds to a neutrino star mass
 $M_{\infty}=1.795\times10^{9}M_{\odot}g_{\nu}^{-1/2}$ for
 a  neutrino mass
 $m_{\nu}=17.2~{\rm keV}/c^2$.
For  a gas of
neutralinos of a mass precisely equal to the neutron mass
 $m_{n}=0.93955~{\rm GeV}/c^{2}$
 and a degeneracy factor
 $g_n=2$,
the infinite density limit
is $M_{\infty}=0.4164~M_{\odot}$, whereas the OV limit
is $M_{\rm{OV}}=0.7091~M_{\odot}$
and $R_{\rm{OV}}=9.1816$ km \cite{HAT}.
Thus, owing to
their compactness,
 neutralino stars could
 mimic the properties of
``machos" which have been detected
in the dark halo of our galaxy,
and which are usually assumed to be baryonic brown dwarfs.
For large $X$, the solutions of the TOV
 equations (\ref{eq:14}) and  (\ref{eq:15})
tend to
\begin{equation}
\mu=\frac{3}{14}\, x \;\;\;{\rm and} \;\;\;
X=\left(\frac{3}{28}\right)^{1/4}x^{-1/2}.
\end{equation}
The pressure and the density thus become
\begin{equation}
P_{\nu}=\frac{c^4}{56\pi}\, \frac{1}{r^2}
\;\;\; {\rm and}\;\;\; \rho_{\nu}=
\frac{3c^{2}}{56\pi}\, \frac{1}{r^2} ,
\end{equation}
yielding the equation of state of radiation
\begin{equation}
P_{\nu}=\frac{1}{3}\, c^{2}\rho_{\nu},
\end{equation}
as expected.
\section{Degenerate neutrino and neutralino matter}
\label{sec3}
We now turn to the discussion of
an astrophysical
 system consisting of degenerate heavy-neutrino
and neutralino matter
that is gravitationally coupled.
As each component satisfies
the equation  of hydrostatic equilibrium separately, i.e.,
Eq.~(\ref{eq:02}) and
\begin{equation}
\frac{dP_{n}}{dr}=-\frac{1}{2}(\rho_{n}c^{2}+P_{n})\frac{d\nu}{dr},
\label{eq:160}
\end{equation}
the total pressure  $P=P_{n}+P_{\nu}$
and the total mass density  $\rho=\rho_{n}+\rho_{\nu}$  will
also obey the same equation
\begin{equation}
\frac{dP}{dr}=-\frac{1}{2}(\rho c^{2}+P)\frac{d\nu}{dr}\  .
\label{eq:100}
\end{equation}
In  addition to the equation of state for neutrino matter,
(\ref{eq:08}) and (\ref{eq:09}), we now have
the equation of state for neutralino matter:
\begin{equation}
P_{n}=K\frac{g_{n}}{g_{\nu}}\left(\frac{m_{n}}{m_{\nu}}\right)^{4}
\left[Y(1+Y^{2})^{1/2}\left(\frac{2}{3}Y^{2}-1\right)
+\log\left(Y+(1+Y^{2})^{1/2}\right)\right],
\label{eq:008}
\end{equation}
\begin{equation}
\rho_{n}=\frac{K}{c^{2}}\frac{g_{n}}{g_{\nu}}\left(\frac{m_{n}}{m_{\nu}}\right)^{4}
\left[Y(1+Y^{2})^{1/2}(2Y^{2}+1)
-\log\left(Y+(1+Y^{2})^{1/2}\right)\right],
\label{eq:009}
\end{equation}
where $g_{n}$ is the spin-degeneracy factor  for
neutralinos and antineutralinos,  and
$Y$
is the local Fermi momentum of neutralino matter (in units of
$m_{n}c$).
Inserting
(\ref{eq:008}) and (\ref{eq:009}) into
(\ref{eq:160}), after integration  we arrive at
\begin{equation}
Y=[(1+Y_{0}^{2}){\rm{e}}^{\nu(0)-\nu(r)}-1]^{1/2}\ ,
\label{eq:170}
\end{equation}
with $Y_{0}=Y(0)$.
Using
(\ref{eq:08}), (\ref{eq:09}), and
the equation of hydrostatic equilibrium (\ref{eq:02}),
a similar relation for the Fermi momentum of neutrinos
(in units of $m_{\nu}c$)
is obtained:
\begin{equation}
X=[(1+X_{0}^{2}){\rm{e}}^{\nu(0)-\nu(r)}-1]^{1/2}.
\label{eq:130}
\end{equation}
Combining (\ref{eq:170})
and (\ref{eq:130}),
the two local Fermi momenta are related by
\begin{equation}
X^2=\frac{(X_{0}^{2}+1)Y^{2}+X_{0}^{2}-Y_{0}^2}{1+Y_{0}^2}.
\label{eq:180}
\end{equation}
The condition
  $X^2 \geq 0$  restricts the range of allowed
  values of $Y$ to
\begin{equation}
Y^{2} \geq \frac{Y_{0}^{2}-X_{0}^{2}}{1+X_{0}^{2}} \ .
\label{eq:190}
\end{equation}
The total pressure and
mass density is given by
\begin{equation}
 P(Y)=P_{n}(Y)+P_{\nu}(X(Y))
\label{eq:400}
\end{equation}
and
\begin{equation}
\rho(Y)=\rho_{n}(Y)+\rho_{\nu}(X(Y)),
\label{eq:401}
\end{equation}
respectively.

We now formulate the coupled differential
equations describing a gravitationally interacting
system of degenerate heavy-neutrino and neutralino
matter.
We first keep the mass of the neutrino halo constant
while varying the mass of the neutralino star.
Introducing the dimensionless
variables $x=r/a_{n}$ and $\mu=m/b_{n}$ with the scales
\begin{equation}
a_{n}=\frac{2}{m_{n}^{2}}\sqrt{\frac{\pi\hbar^3}{g_{n}cG}}
\;\;\; {\rm and} \; \; \;
 b_{n}=\frac{2}{m_{n}^{2}}\sqrt{\frac{\pi\hbar^{3}c^3}{g_{n}G^3}},
\label{eq:ab}
\end{equation}
the relevant
 TOV equations can be written
in the form
\begin{eqnarray}
\frac{dY}{dx}&=&-\frac{1+Y^2}{Y(x^{2}-2\mu x)}
\left\{\mu+x^{3}\left[Y(1+Y^{2})^{1/2}
\left(\frac{2}{3}Y^{2}-1\right)\right.\right.
+\log\left(Y+(1+Y^{2})^{1/2}\right)
\nonumber \\
&&
+\left(\frac{m_{\nu}}{m_{n}}\right)^{4}\frac{g_{\nu}}{g_{n}}
\left.\left. \left(X(1+X^{2})^{1/2}\left(\frac{2}{3}X^{2}-1\right)
+ \log\left(X+(1+X^{2})^{1/2}\right)\right)\right]\right\},
\label{eq:70}
\end{eqnarray}
\begin{eqnarray}
\frac{d\mu}{dx}&=&x^{2}\left\{Y(1+Y^{2})^{1/2}(2Y^{2}+1)+
\log\left(Y+(1+Y^{2})^{1/2}\right)\right. \nonumber\\
& & +\left(\frac{m_{\nu}}{m_{n}}\right)^{4}\frac{g_{\nu}}{g_{n}}
\left.\left[X(1+X^{2})^{1/2}(2X^{2}+1)
 +\log\left(X+(1+X^{2})^{1/2}\right)\right]\right\}\  ,
\label{eq:71}
\end{eqnarray}
where  $X$ is
related to $Y$ through (\ref{eq:180}).
If the condition (\ref{eq:190}) is not fulfilled, i.e., the
neutrino pressure and density have already vanished,
the  system is solved with the $Y$ terms
 describing the neutralinos only.

In order to solve
Eqs.~(\ref{eq:70})
and (\ref{eq:71})
numerically,
we fix the Fermi momentum of neutrinos
(in units of $m_{\nu}c$)
at the center
 and vary the central values of
the corresponding quantity $Y_{0}$ for
neutralinos.
The total mass (including neutrinos
and neutralinos) enclosed within the radius
 $R_{n}$
 of the neutralino star
 is shown in Fig.~\ref{fig4}.
 Here, the
 neutrino mass and the degeneracy factor are
 taken to be $m_{\nu}=17.2~{\rm keV}/c^{2}$
 and $g_{\nu}=2$,
 respectively,
 while for
the neutralino mass we have chosen
 $m_{n}=939.55~{\rm MeV}/c^{2}$
 and $g_n=2$, with the scales
 $a_{n}=6.8304~{\rm km}$
and $b_{n}=4.6257~M_{\odot}$.
For small
 neutralino-star masses, the total mass
 enclosed in $R_n$
  scales as $R_n^{3}$,
  corresponding to a constant density
  governed by the gravitational potential of
  the surrounding supermassive neutrino halo.
  However, as the radius of the neutralino star
approaches that of a ``free" neutralino star,
the gravitational potential of the neutralino star
becomes dominant and the mass now scales
as
$R_n^{-3}$ up to the OV limit.
Thus there is always a maximal radius of a
neutralino star
within a neutrino halo of a given mass.
Substituting   neutralinos by neutrons,
we must take care of the
fact that
(i) the neutron interacts strongly in the nuclear
medium
(simulated, e.g., by an effective mass) and
(ii) the neutron decays through weak interactions.
Thus,  stable neutron stars
can exist only in the
 range  from
$0.2~M_{\odot}$ to $2~M_{\odot}$ \cite{TSUR}, where the binding
energy is larger than the Q value for the neutron decay.

It is instructive to study the properties
of a degenerate
  gas of neutralinos and
neutrinos  in the nonrelativistic
approximation.
In the limits
 $X\ll 1$ and $Y\ll 1$,
 (\ref{eq:70}) and (\ref{eq:71})
 simplify to
\begin{equation}
\frac{dY}{dx}=-\frac{\mu}{x^{2}Y},
\end{equation}
\begin{equation}
\frac{d\mu}{dx}=\frac{8}{3}x^{2}[Y^{3}+\frac{g_{\nu}}{g_{n}}
\left(\frac{m_{\nu}}{m_{n}}\right)^{4}(Y^{2}+X_{0}^{2}-Y_{0}^{2})^{3/2}]\ ,
\label{eq:200}
\end{equation}
with the  boundary conditions
\begin{equation}
 \mu(0)=0 ;\;\;\;\; Y^{2}\ge  Y_{0}^{2}-X_{0}^2;\;\;\;\;  Y(0)=Y_{0}\ .
\end{equation}
This  system  of equations can be rewritten
 in the form of a
Lan\'{e}-Emden type equation by introducing $\Theta_{n}=Y^{2}$,
$\Theta_{\nu}=X^{2}$, and a new radial dimensionless
radial variable $\xi=4x/\sqrt{3}$
\begin{equation}
\frac{1}{\xi^2}\frac{d}{d\xi}\left(\xi^{2}\frac{d\Theta_{n}}{d\xi}\right)
=-[\Theta_{n}^{3/2}+\frac{g_{\nu}}{g_{n}}
\left(\frac{m_{\nu}}{m_{n}}\right)^{4}(\Theta_{n}+\Theta_{\nu 0}-
\Theta_{n0})^{3/2}]\ ,
\end{equation}
where $\Theta_{n0}$ and $\Theta_{\nu 0}$ are the central values
of the neutralino and neutrino densities, respectively.
For very small neutralino densities,
i.e., $Y \ll 1$ and $Y_{0} \ll 1$,
the mass equation (\ref{eq:200}) can be integrated to give
\begin{equation}
\mu(x)=\frac{8}{9}\left(\frac{m_{\nu}}{m_{n}}\right)^{4}
X_{0}^{3}x^3 \ ,
\end{equation}
which confirms the conclusion drawn
in the context of Fig.~\ref{fig4}.

We now turn to the case of a
neutralino star of constant mass surrounded by
a neutrino halo of variable mass.
The TOV equations
 written in terms of the functions $X$ and $\mu$
 may be obtained from
 (\ref{eq:70}) and (\ref{eq:71}),
 in which we make the replacements
 $X \leftrightarrow Y$,
 $g_{\nu}\leftrightarrow g_{n}$,
 and $m_{\nu}\leftrightarrow m_{n}$.
 Thus, we find
\begin{eqnarray}
\frac{dX}{dx}&=&-\frac{1+X^2}{X(x^{2}-2\mu x)}
\left\{\mu+x^{3}\left[X(1+X^{2})^{1/2}
\left(\frac{2}{3}X^{2}-1\right)\right.\right.
+\log\left(X+(1+X^{2})^{1/2}\right)
\nonumber \\
&&
+\left(\frac{m_{n}}{m_{\nu}}\right)^{4}\frac{g_{n}}{g_{\nu}}
\left.\left. \left(Y(1+Y^{2})^{1/2}\left(\frac{2}{3}Y^{2}-1\right)
+ \log\left(Y+(1+Y^{2})^{1/2}\right)\right)\right]\right\},
\label{eq:250}
\end{eqnarray}
\begin{eqnarray}
\frac{d\mu}{dx}&=&x^{2}\left\{X(1+X^{2})^{1/2}(2X^{2}+1)+
\log\left(X+(1+X^{2})^{1/2}\right)\right. \nonumber\\
& & +\left(\frac{m_{n}}{m_{\nu}}\right)^{4}\frac{g_{n}}{g_{\nu}}
\left.\left[Y(1+Y^{2})^{1/2}(2Y^{2}+1)
 +\log\left(Y+(1+Y^{2})^{1/2}\right)\right]\right\}\  ,
\label{eq:251}
\end{eqnarray}
with $X$ and $Y$ subject to the condition
\begin{equation}
X^{2} \geq \frac{X_{0}^{2}-Y_{0}^{2}}{1+Y_{0}^{2}}\ .
\label{eq:252}
\end{equation}
If this condition is not satisfied, i.e.,
the pressure and  density
of neutralinos have already vanished,
 (\ref{eq:250}) and
(\ref{eq:251})
are solved without the $Y$ terms, i.e., for neutrinos
only.
Choosing the OV limit as
 the mass of the neutralino star,
 i.e.,
$M_{\rm OV}^n=0.7091~M_{\odot}$
for $m_n=0.93955$ GeV/$c^2$
and $g_n=2$,
and varying the central Fermi momentum
$X_0$,
one can find the total mass
(including
 neutralinos
and neutrinos) as a function of the radius $R_{\nu}$ of
the neutrino halo.
This
scenario is reflected in  Fig.~\ref{fig5} where the length
and mass scales are
$a_{\nu} =2.0381\times 10^{10}~{\rm km}$ and $b_{\nu}
=1.3803\times10^{10}~M_{\odot}$,
respectively.
Here the neutrino mass and  the degeneracy
factor have been chosen as
 $m_{\nu}=17.2~{\rm keV}/c^{2}$
 and $g_{\nu}=2$, respectively.
At the turning point $A$,
the total mass enclosed within the radius
$R_A=R_{\rm OV}^n= 9.1816~{\rm km}$
of the neutrino halo
is $M_A=M_{\rm OV}^n=0.7091~M_{\odot}$.
At the turning point $B$, the total mass
enclosed within the radius
 $R_B=0.9912~{\rm pc}$
 of the neutrino halo is
$M_B=3.3453~M_{\odot}$.
It is interesting to note that,
also in this case,
there is a maximal radius $R_B$
of the neutrino halo, for a given mass of
the neutralino star.

Replacing the neutralino star by a
baryonic star, such as a
neutron star, a white dwarf, or an ordinary star,
the only thing that will change in Fig.~\ref{fig5}
is the point $A$ at which the enclosed mass
starts deviating from the constant value, which depends,
of course,on the mass $M_n$ of the central object.
Thus for
$M_n \stackrel{\textstyle >}{\sim} M_{\odot}$,
the halo will have a size of a few light-years
and a mass of a few times that of the central
baryonic or nonbaryonic star.
In this context, it is important to note that
if every baryonic star is surrounded by such
a neutrino halo, the degeneracy pressure
of the neutrino halo would prevent
stars from approaching each other
closer than a distance of a few
light-years.
In such a scenario,
a large fraction of  galactic dark matter would be
nonbaryonic.
A further attractive feature
of this scenario is that a neutrino mass of the order
of 14 or 15 keV/$c^2$,
could, at the same time,
set the mass scale of the supermassive compact
dark objects at the centers of galaxies
and the scale of  interstellar distances
in galaxies.

To investigate the consequences of this
idea in more detail,
let us assume that the sun is surrounded by a degenerate
neutrino halo.
In the vicinity of the sun,
in the region of the size of the planetary
system,
the neutrino density is governed by the
gravitational potential of the sun.
In fact, the mass due to neutrinos contained
within a radius $r$ is,
 in the vicinity of a baryonic or
a nonbaryonic star of mass $M_n$,
given in the nonrelativistic approximation
\cite{RDV,VILT}
by
\begin{equation}
\frac{M_{\nu}}{M_{\odot}}=
1.34\times 10^8 g_{\nu}
\left(\frac{M_{\nu}}{M_{\odot}}\right)^{3/2}
\left(\frac{m_{\nu}c^2}{17.2{\rm keV}}\right)^4
\left(\frac{r}{\rm AU}\right)^{3/2} ,
\label{eq:1000}
\end{equation}
where AU $=1.496\times 10^8$ km is the
astronomical unit.
This means that for
$m_{\nu}c^2=17.2$ keV,
$g_{\nu}=2$, and $M_n=M_{\odot}$,
the  mass of the neutrino
(and antineutrino) halo contained
within the earth's orbit
would be
 $M_{\nu}=2.68\times 10^{-8} M_{\odot}$.

From the Pioneer 10 and 11 and the Voyager 1 and 2
ranging data \cite{and}
we know that the dark mass contained within
Jupiter's orbit is
$M_d =(0.12\pm 0.027) \times 10^{-6} M_{\odot}$
and within Neptun's orbit
$M_d \leq 3\times 10^{-6} M_{\odot}$.
Of course,
the Jupiter data should be taken only
as a lower limit,
as Jupiter tends to eject almost any matter
within its orbit
\cite{and}.
Nevertheless,
taking the Jupiter data at face value,
and interpreting dark matter as
degenerate neutrino matter,
the neutrino mass limits are
\begin{eqnarray}
 12.6\,{\rm keV}/c^2
 \leq &m_{\nu}&  \leq
 14.2\,{\rm keV}/c^2
 \;\;\;\;\; {\rm for}
\;      g_{\nu}=2,
\nonumber \\
 10.6\,{\rm keV}/c^2
 \leq &m_{\nu}&  \leq
 12.0\,{\rm keV}/c^2
 \;\;\;\;\; {\rm for}
\;      g_{\nu}=4.
\end{eqnarray}
 For dark matter within Neptun's
 orbit, the neutrino mass limits are
\begin{eqnarray}
  &m_{\nu}&  \leq
 15.6\,{\rm keV}/c^2
 \;\;\;\;\; {\rm for}
\;      g_{\nu}=2,
\nonumber \\
 &m_{\nu}&  \leq
 13.1\,{\rm keV}/c^2
 \;\;\;\;\; {\rm for}
\;      g_{\nu}=4.
\label{eq:1300}
\end{eqnarray}
In summary, considering
 (\ref{eq:1100}),
 (\ref{eq:1200}),
 and (\ref{eq:1300}),
 a neutrino mass-range
\begin{eqnarray}
 14.3\,{\rm keV}/c^2
 \leq &m_{\nu}&  \leq
 15.6\,{\rm keV}/c^2
 \;\;\;\;\; {\rm for}
\;      g_{\nu}=2,
\nonumber \\
 12.0\,{\rm keV}/c^2
 \leq &m_{\nu}&  \leq
 13.1\,{\rm keV}/c^2
 \;\;\;\;\; {\rm for}
\;      g_{\nu}=4,
\end{eqnarray}
seems to be consistent with all reliable data.
\section{Conclusions}
\label{sec4}
 We have studied   degenerate fermion
 stars, consisting of massive neutrinos or neutralinos,
 or both. We have shown that the
   existence of such objects may have important
   astrophysical implications.

For neutrino masses  in the  range of several keV,
neutrino stars are natural candidates for the supermassive
dark objects at the centers of galaxies.
Assuming that the most massive
 object,
such as the compact dark object at the center of M87,
is a neutrino star at the OV limit,
the neutrino mass required for this scenario
should be between
  $10~{\rm keV}/c^{2}$  and
  $16~{\rm keV}/c^{2}$,
  depending on the degeneracy factor $g_{\nu}$.

Furthermore,  interpreting
 the supermassive dark object at the center of
our galaxy as a neutrino star,
we obtain
from the upper limit of the size of this object,
a lower bound on
 the neutrino mass
which overlaps with the range mentioned above.
In addition,
our interpretation explains the
so-called ``blackness problem" of Sgr A$^*$
in a natural way.

By studying a two-component system
consisting of neutralinos in the GeV mass range
and neutrinos in the keV mass range, we have found
that there is always a maximal mass and radius of a neutralino star
within a neutrino halo of a given mass.
Owing to their compactness, neutralino stars could mimic the
properties of ``machos".

Finally, assuming that ordinary stars are
surrounded by  degenerate neutrino halos
of maximal size,
 a neutrino mass of
the order of 15 keV/$c^2$
sets the scale of interstellar distances to a few
light-years.

\acknowledgments

This research was supported by the Foundation
for Fundamental Research (FFR).
F.M. gratefully acknowledges financial support by the
Deutscher Akademischer Austauschdienst.


\newpage
\begin{figure}
\caption{The total mass $M$ of a
 neutrino star as a function of its radius $R$.}
\label{fig1}
\end{figure}

\begin{figure}
\caption{The total mass $M$ of a neutrino star as a function
 of its total number of particles $N$.}
\label{fig2}
\end{figure}

\begin{figure}
\caption{The relative mass defect  $\Delta$
 as a function of  the radius $R$ of the neutrino star.}
\label{fig3}
\end{figure}

\begin{figure}
\caption{The total mass (including neutralinos and neutrinos)
enclosed within the radius $R_{n}$ of the neutralino star
for various masses $M_{\nu}$ of the neutrino halo.}
\label{fig4}
\end{figure}

\begin{figure}
\caption{The total mass of neutralinos  and
neutrinos contained within the radius $R_{\nu}$
 of the neutrino halo around a neutralino star.}
\label{fig5}
\end{figure}
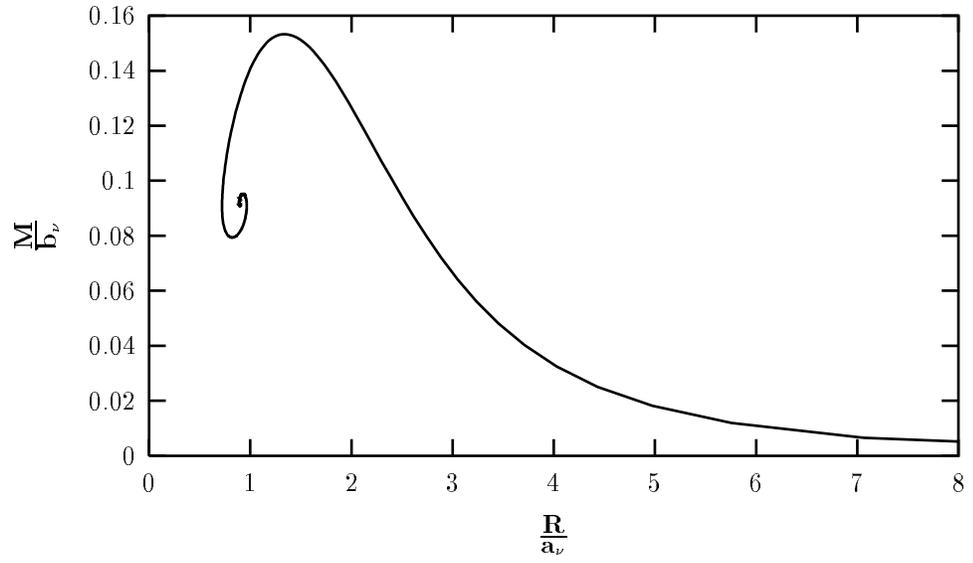
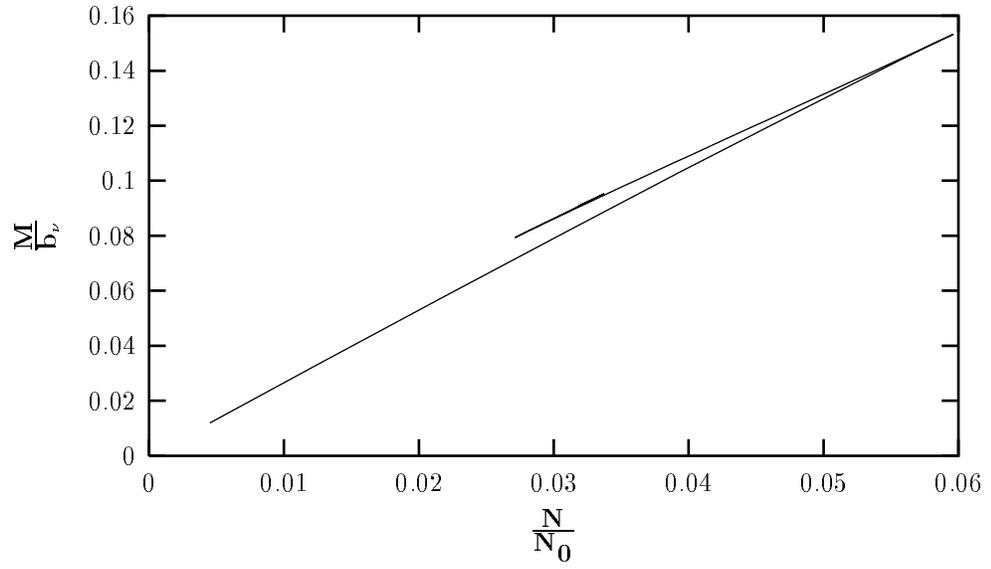
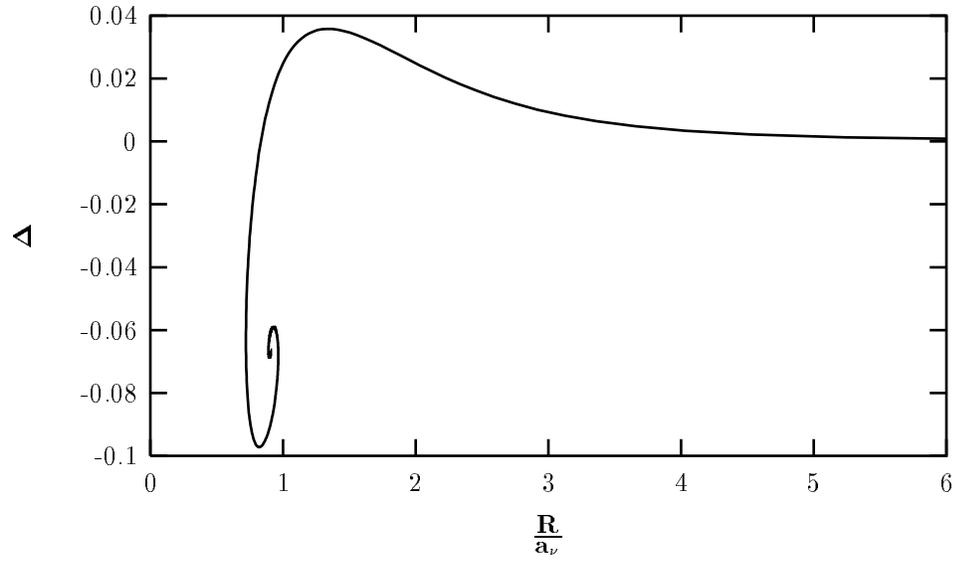
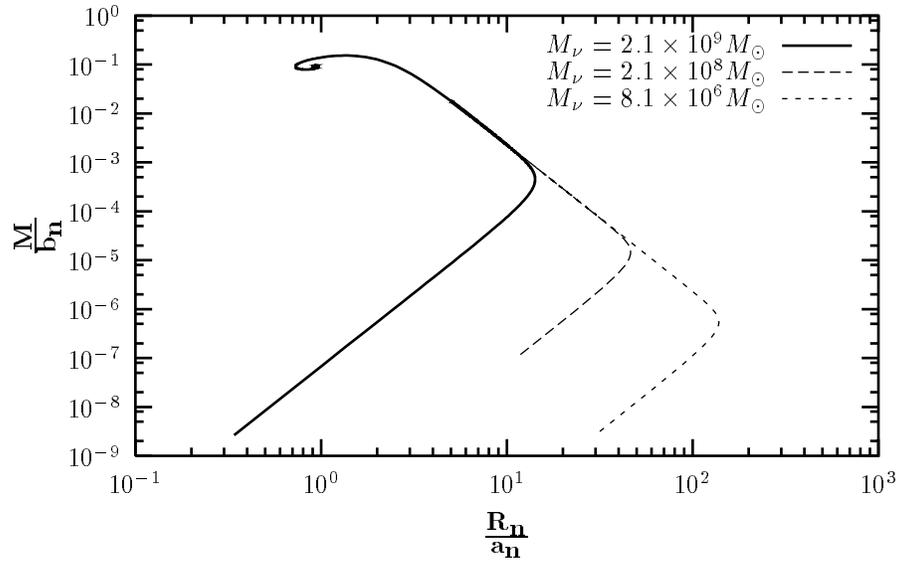
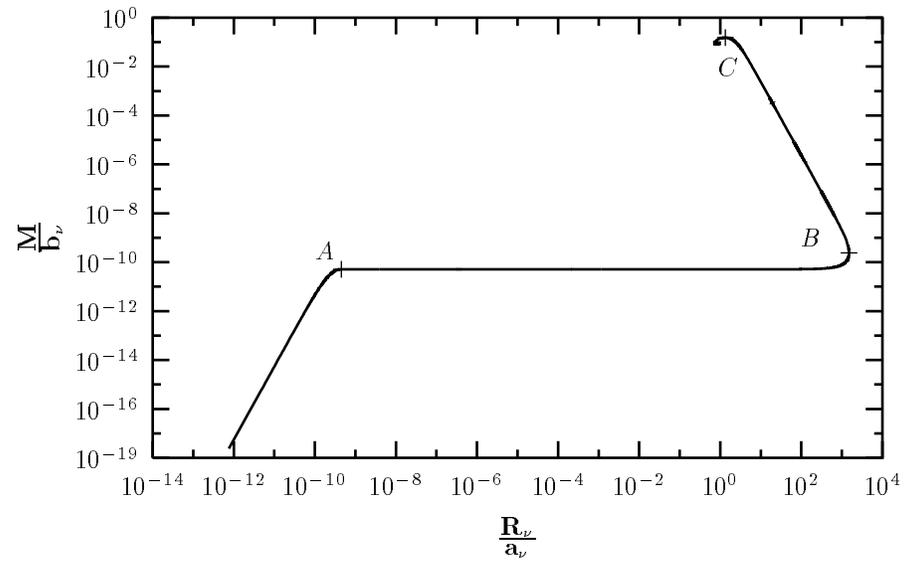
\end{document}